Stanislav Dolgopolov
dolgopolov-s@list.ru

# Superconductivity in Nanostructures as a Consequence of Local Pairing and Bose-Einstein-Condensation of Pairs

**Abstract**: The comment to article "*Coexistence of Diamagnetism and Vanishingly Small Electrical Resistance at Ambient Temperature and Pressure in Nanostructures*" [1] shows the role of local electron pairing and Bose-Einstein-Condensation of pairs for superconductivity at $T_c$ = 286 K in nanostructures.

The article "*Coexistence of Diamagnetism and Vanishingly Small Electrical Resistance at Ambient Temperature and Pressure in Nanostructures*" [1] is an informative result, pointing to the local electron pairing and Bose-Einstein-Condensation (BEC) of pairs as bosons.

In the research $T_c$ is tuned by **the disorder density and defect nature** in nanostructure (NS), containing 3 fractions: Au, Ag, unknown insulating fraction. The disorder causes consequences A and B:

**A.** The carrier trapping, reducing the free electron density in respect to the density in pure metals. So the Fermi level ($E_F$) of NS is lower than $E_F$ of pure metals and may be close to a Brillouin zone edge, where electronic standing waves occur. The standing waves are local and have zero-momenta, so they can form singlet pairs despite a huge kinetic energy on the Fermi surface [2].

$E_F$ (Ag) ≈ $E_F$ (Au) ≈ 5.5 eV, the highest Brillouin zone edge is in direction [110] = $h^2/(8m_e R_{110}^2)$=4.5 eV, where $R_{110}$ is corresponding lattice parameter. Thus the trapping and the insulating fraction should reduce $E_F$ by only 1 eV (plausible value for nano-disordered metals).

**B.** The disorder provides new local electron states on the Fermi surface; this locality (like standing waves) reduces the length of electronic wave packets, what enhances the singlet bond of electrons (exchange energy is strongest locally, like in atoms). Thus the shorter the localization length, the higher the pairing temperature ($T^*$).

The result reasonably combines the locality with the **Bose-Einstein Condensation**. Using the link between the temperature of Bose-Einstein Condensation ($T_{BEC}$) and bosonic density ($N_b$), we can estimate the length of electron localization ($L$):

$L \leq (N_b)^{-1/3}$    (1)

$L \leq [3.31 \cdot \hbar^2/(T_{BEC} \cdot 2 \cdot m_e \cdot k_B)]^{0.5}$    (2)

At $T_{BEC}$= 286 K we obtain $L \leq 2.3$ nm. At FIG. S2 in [1] the average distance between neighboring defects is ≈ 2 nm. Assuming that each localization is pinned to one or a few defect points, the BEC at 286 K is possible. So $T_c$ grows with the decreasing Au-fraction, since the disorder density is inversely related to the Au-fraction (FIG.5a in [1]). The final BE-condensate is not local, since the BEC is a macroscopic coherent state of all pairs.

The disorder density is inhomogeneous in volume, causing a variation range of $T_{BEC}$, since $T_{BEC}$ is related to the disorder density. The BEC - process starts at $T_c$. A further decreasing $T$ gives rise to BEC in new regions with lower $T_{BEC}$, so the superfluid density depends on $T$, a certain $T$ is linked to a certain superfluid density. This link is independent of the $T$ - run in one (the same) sample, since $T$ doesn't affect the disorder density, so the sample **noise may be repeatable** on the $T$-scale.

In pure metals at 300 K is impossible to combine the locality and a high Fermi level (a few eV), therefore $T_c$ is low. The insulating effect in NS plays a crucial role, since it strongly affects the localization.

A possible localization mechanism of conduction electrons in the Ag-Au nanostructure is a **reflection of electrons from Ag-sphere-boundaries**. This reflection may be sufficiently strong, if the Ag-sphere is (partially or fully) covered by an insulating layer. So far the nature of the insulating fraction in the Ag-Au nanostructure is unknown, however, its presence is probable, since some samples are insulating.

The electron reflection and localization in Ag-spheres may be also related with the Volta potential between Au and Ag nanoclusters. The work function in Au (>4.6 eV) is stronger than in Ag (≈ 4.3 eV); what can lead to electron depletion within the Ag-sphere, so the Ag-sphere becomes positively charged and the surrounding Au becomes negative. Thus a stronger work function in Au (than in Ag) puts the Ag-sphere into a 3D potential well, so a multiple reflection of Ag-electrons can occur on the well boundaries. On a Brillouin zone edge, these multiply reflected electrons become standing waves with zero-momenta, enabling the pairing. Running waves (even within a cluster) cannot pair up because of a huge kinetic energy on the Fermi surface. We note, the exchange energy of electrons and, thus, their pairing temperature $T^*$ are directly related to the depth of 3D potential well [2].

Localization factors (Volta potential, impurities, disorder) can work together, enhancing each other for the electron localization and for $E_F$ reduction. Thus the localization occurs within an Ag-sphere, the localization length is roughly equal to the Ag-sphere size ($\approx$ 1nm).

Note: the positively charged Ag - cluster has a lower electron density and, thus, a lower $E_F$. Obviously, this $E_F$ reduction in the Ag-cluster depends on the cluster size; the smaller the Ag-cluster, the stronger the $E_F$ reduction, since $E_F$ in a very large cluster tends to $E_F$ in bulk ($\approx$ 5.5 eV). Therefore there is an optimal Ag-cluster size, where $E_F$ is leaned on the Brillouin zone edge (i.e. $E_F \approx$ 4.5 eV), so standing waves within the Ag-sphere can occur, giving rise to the local pairing. Thus the superconductivity is tuned by the $E_F$ tuning. This tuning is universal for superconductors; doping, pressure, gate voltage, film thickness etc. are parameters, which shift $E_F$ in respect to the zone edges. If $E_F$ crosses a zone edge, then a $T_c$ dome is observable [2].

Work functions of metals vary depending on the direction in crystal, [100], [110] etc. Hence the strength of localization and $E_F$ depend on the alignment of Ag - cluster in respect to Au - cluster. Probably, this alignment is challenging to reproduce in all samples, so $T_c$ is also difficult to reproduce.

Every pair emits a pairing energy $2|\Delta|$ and its associated momentum $p$, which should be added to the initial zero-momentum of the pair. The final momentum of the pair $p$ is:

$$p \approx (2|\Delta| \cdot 2 \cdot 2m_e)^{0.5} \quad (3)$$

Then the wave length of the bosonic pair $\lambda_b$ is:

$$\lambda_b = h/p \quad (4)$$

Assuming that $2|\Delta| \approx 3.5 \cdot k_B \cdot T_c$ and using the equations (3) and (4), we can find $\lambda_b$ for $T_c$ = 286 K:
$\lambda_b$ (286 K) $\approx$ 3 nm .

This result indicates that **the condensed pairs are not local**; the average distance between Ag-clusters is $\approx$ 1.9 nm (calculated for Ag-cluster size 1 nm and Ag mole fraction $\approx$ 15 %), so a boson with a wave length 3 nm can hop between neighboring Ag-spheres. It is in line with the BEC of bosonic pairs; as shown above the BEC at $T$=286 K is possible if the average distance between pairs is roughly 2 nm.

Thus $T_{BEC}$ depends on the distance between Ag-spheres, whereas the localization length and its associated pairing temperature $T^*$ depend on the Ag-sphere size, so $T_{BEC}$ is not necessarily equal to $T^*$. $T_c$ is the lowest value from $T_{BEC}$ and $T^*$. On FIG.5a in [1] $T_c$ grows with the decreasing Au-fraction; hence $T_c$ grows with the decreasing distance between Ag-spheres, indicating that $T_c = T_{BEC} < T^*$.

Probably, the fast sample degradation is related with the degradation of the reflecting layer around Ag. Therefore, the reproducibility of $T_c$ is extremely difficult.